\documentclass[aps,prb,showpacs,twocolumn]{revtex4}  
\usepackage{graphicx}
\usepackage{natbib}
\usepackage{color}
\usepackage{amsmath,amssymb,subfigure}                         
\def\gapx{\lower 2pt \hbox{$\buildrel>\over{\scriptstyle{\sim}}$\ }}
\def\lapx{\lower 2pt \hbox{$\buildrel<\over{\scriptstyle{\sim}}$\ }}
\def\he4{$^4$He}
\def\paraH2{{\it p}-H$_2$}
\def\orthoD2{{\it o}-D$_2$}
\def\Am2{\AA$^{-2}$}

\begin{document}

\title{Molecular hydrogen isotopes adsorbed on krypton-preplated
\\ graphite: Quantum Monte Carlo simulations}
\author{Joseph Turnbull and Massimo Boninsegni} 
\affiliation{Department of Physics, University of Alberta, Edmonton, 
    Alberta, Canada T6G 2J1}
\date{\today}

\begin{abstract}  

Adsorption of \textit{ortho}-deuterium and \textit{para}-hydrogen films on a graphite substrate, pre-plated with a single atomic layer of krypton, is studied theoretically by means of quantum Monte Carlo simulations at low temperature.  Our model explicitly includes substrate corrugation. Energetic and structural properties of these adsorbed films are computed for a range of hydrogen coverages.  Thermodynamically stable adsorbed films are solid, with no clear evidence of any liquid-like phase.  Quantum exchanges of \textit{ortho}-deuterium and \textit{para}-hydrogen are essentially absent in this system, down to zero temperature; consequently, this system displays no superfluidity in this limit. Our simulations provide evidence of a stable domain wall fluid at low temperature, consistently with recent experimental observations.

\end{abstract}

\maketitle

\section{INTRODUCTION}

Studies of low temperature adsorption of highly quantal fluids, such
as helium or molecular hydrogen are 
motivated by the fascinating properties
that such adsorbed quantum films display, often considerably different
than those of the bulk.  Fluids of {\it ortho}-deuterium (\orthoD2) and {\it para}-hydrogen (\paraH2) molecules are interesting physical systems, because these molecules have mass equal to (in the case of \orthoD2) or half (in the case of \paraH2) the mass of a helium atom, rendering zero-point motion significant at low temperature.  In addition, each molecule is
a composite boson, and therefore it is conceivable that, at low enough temperature,
a hydrogenic fluid might display physical behavior similar to that of liquid
helium, including superfluidity (SF).\cite{ginzburg72}

However, unlike helium, bulk hydrogen solidifies at low temperature 
($T_{\rm c} \approx$ 14 K); this solidification has so far prevented the observation of phenomena such as Bose Einstein condensation (BEC) and SF which, in the bulk, are speculated
to occur in the liquid phase below $T$ $\approx$ 6 K.  Solidification occurs because of the depth of the 
attractive well of the potential between two hydrogen molecules, which is significantly greater than that between two helium atoms. Several attempts have been made\cite{bretz81,maris86,maris87,schindler96} to supercool bulk liquid \paraH2, but the search for SF (in the bulk) has so far not met with success.

Toward the aim of maintaining a stable liquid phase of molecular hydrogen down to low enough temperatures for SF to be observed, several avenues can be explored, including reduction of dimensionality.  The predicted\cite{boninsegni04b} theoretical value of the melting temperature  $T_{\rm m}$ for a solid film of  \paraH2  in two dimensions (2D), is slightly less than 7 K, i.e., approximately half that of three-dimensional bulk \paraH2, but still significantly higher than that at which the system, if it remained a liquid, would undergo Bose Condensation and turn superfluid, estimated at $\sim$ 2 K in 2D. However,  in an actual experimental realization of a quasi-2D system, i.e.,  a film of hydrogen molecules adsorbed on a substrate, quantum zero-point motion of adsorbed particles in the direction perpendicular to the substrate can be significant, as calculations for adsorbed helium films on alkali metal substrates have shown.\cite{boninsegni99} One might speculate 
 that such zero-point motion may result in an effective screening of the hydrogen intermolecular interactions, possibly leading to a reduction of $T_{\rm m}$, with respect to the purely two-dimensional case. 

With this aim in mind, extensive experimental studies of hydrogen films adsorbed on a different substrates have been carried out, using various techniques. \cite{nielsen80,lauter90,wiechert91,vilches92,cheng93b,mistura94,ross98}  One of the most remarkable findings, is that the melting  temperature $T_m$ of a solid \paraH2 monolayer\cite{vilches92}  can indeed be significantly less than that of  purely 2D \paraH2. The interesting question is whether by ``fine tuning" of the substrate, $T_{\rm m}$ may be lowered to the point where interesting, collective quantum many-body phenomena could become observable, in some thermodynamically stable liquid phase. 

In a recent neutron scattering investigation of \orthoD2 films adsorbed on a krypton pre-plated graphite substrate,\cite{wiechert04} evidence of a stable ``liquid-like'' phase of  \orthoD2 down to $T\sim$ 1.5 K was reported (termed a ``domain-wall'' fluid by the authors).  This result is obviously of great interest, as it appears that the substrate considered in this work may indeed provide an avenue to the observation of  the long sought superfluid phase of molecular hydrogen, given the low temperature down to which a liquid-like phase may exist. Motivated by this experiment, we have undertaken a theoretical study of the low temperature phase diagram of \paraH2 and \orthoD2 films adsorbed on such a substrate.  

In a previous study,\cite{turnbull05} we determined that \paraH2, adsorbed upon a substrate consisting of graphite preplated with a \textit{commensurate} monolayer of krypton forms two thermodynamically stable monolayer phases at low $T$, both {\it solid}; one is commensurate with the krypton layer, 
the other incommensurate.  No evidence was observed of a liquid phase at $T=0$ for these or any intermediate coverages, and quantum exchanges of \paraH2 were found to be greatly suppressed. Quantum zero-point motion in the direction perpendicular to the substrate was found to be less significant than that observed for \paraH2 adsorbed on lithium,\cite{boninsegni04_li} from which one can conclude that this mechanism ought not yield a significant reduction in the melting temperature. In that study, while the Kr layer was modeled explicitly, the underlying graphite substrate was assumed smooth.   

In this paper, we again explore this problem, but with two main modifications with respect to Ref. 
\onlinecite{turnbull05}, namely {\it a}) we assume this time a \textit{compressed} krypton layer preplating the graphite, i.e., {\it incommensurate} with the underlying substrate lattice structure, and {\it b}) we {\it explicitly} model corrugation of the underlying graphite substrate, in some of our simulations. Furthermore, in order to allow for a direct comparison with Ref. \onlinecite{wiechert04}, we study adsorption of  both \paraH2 and \orthoD2.   
Energetic, structural and superfluid properties are investigated using Path Integral ground state (PIGS) Monte Carlo simulations, as well as the continuous-space {worm algorithm}\cite{boninsegni06worm1,boninsegni06worm2} for simulations at finite temperature.  

The main results of this study are the following:
\begin{enumerate}
\item{No evidence is observed of a thermodynamically stable {liquid} phase in \orthoD2 or \paraH2.
In all cases, equilibrium phases are solid monolayers, commensurate with the underlying 
krypton plating, with coverage (i.e., 2D density) $\theta_\circ$=0.07253 \Am2 for both isotopes.} 
\item {An \orthoD2 monolayer composed of commensurate domains separated by domain walls is found to be thermodynamically stable, compressible 

in a relatively extended range of coverages.}
\item{Quantum exchanges of \paraH2 or \orthoD2 molecules are essentially absent in the $T\to 0$ limit;  consistently, no evidence of a finite  superfluid response of either isotope is ever observed.}
\item{Our results are consistent with the presence of a ``domain-wall fluid'', as suggested in Ref. \onlinecite{wiechert04}.  Though no dynamical information can be reliably extracted using the QMC techniques employed here, visual inspection of  \textit{instantaneous configurations} generated by our sampling procedure suggests that such a domain-wall fluid indeed form; however, such a fluid is classical in character, and does not support SF of either hydrogen isotope even at $T$=0. } 
\end{enumerate}

The remainder of this manuscript is organized as follows:  Sec. \ref{model} offers a description of the model used for our system
of interest, including a discussion of the potentials and the justifications for the main underlying 
assumptions.  Sec. \ref{method} involves a brief discussion of the computational techniques and specific details of implementation, in addition to details of calibration and optimization.  The results are presented in Sec. \ref{results}; finally, Sec. \ref{conclusions} is a summary of the 
findings and our concluding remarks.

\section{MODEL}
\label{model}

We consider a system of $N$ hydrogen molecules, regarded as point particles of nuclear spin zero (i.e., obeying Bose statistics; this discussion is entirely equivalent for both isotopes).  Molecules sit above a substrate consisting of a single atomic layer of krypton, below which is a graphite substrate. The Kr layer consists of  ($L$) point-like atoms,  pinned at fixed positions ${\bf R}_k$ ($k=1,2,...,L)$ (an assumption justified by their relatively large mass).  They are arranged in a perfectly two-dimensional triangular lattice, with a spacing\cite{wiechert04} of 3.99 \AA\, and sit at a height of 3.46 \AA\ over the top layer of C atoms of the graphite substrate; this latter distance corresponds to the energetic minimum of the most accurate Kr-graphite potential available,\cite{steele,gooding} if graphite is regarded as smooth.

If a smooth graphite substrate is assumed, the model quantum many-body Hamiltonian is therefore as follows: 
\begin{eqnarray}\label{hm}
\hat{H}&=&-\frac{\hbar^{2}}{2m}\sum_{i=1}^{N}\nabla_{i}^{2}  +
\sum_{i<j}V(r_{ij})  \nonumber \\
&+& \sum_{i=1}^N\sum_{k=1}^{L} U (|{\bf r}_i-{\bf R}_k|)  +
\sum_{i=1}^N {\tilde U}({z}_i) 
\end{eqnarray}

Here, $m$ is the mass of a hydrogen molecule, 
$\{{\bf r}_i\}$ 
(with $i$=1,2,...,$N$) are the positions of the hydrogen molecules, $r_{ij}\equiv |{\bf r}_i-{\bf r}_j|$; $z_{i}$ is the height of the $i$th hydrogen molecule above the graphite surface. 
$V$ is the potential describing the interaction between any two hydrogen molecules, and
$U$ represents the interaction of a hydrogen molecule with a Kr atom.  Finally, $\tilde U$ represents the interaction of a hydrogen molecule with the smooth graphite substrate.

As we are investigating \paraH2 and \orthoD2, which have nearly spherically symmetric electron densities, all pair potentials are assumed to depend only on relative distances.  
The interaction  $V$ is described by the Silvera-Goldman potential,\cite{Silvera1} which provides an accurate description of energetic and 
structural properties of condensed \paraH2 and \orthoD2.\cite{johnson96,operetto} 

The interaction of a hydrogen molecule and a Kr atom is modeled using a 
standard 6-12 Lennard-Jones (LJ) potential; we make use of the Lorentz-Berthelot mixing rule,\cite{wang97,macgowan86} yielding $\epsilon=75.6$ K and 
$\sigma=3.3$ \AA, for our purposes consistent with current state-of-the-art potential energy surfaces (for example, numerical agreement is within $\approx$ 5\% for the well depth).\cite{LeRoy05}  
The Kr atomic monolayer is uniform, with each Kr atom kept fixed in its lattice position for the duration of the simulation.  

We use a simple ``3-9" potential to describe the interaction of hydrogen molecules with the smooth graphite substrate,\cite{gatica04}  i.e.,
\begin{eqnarray}\label{sim}
\tilde U(z_{i}) = \frac{4 C^{3}}{27 D^{2} z^{9}} - \frac{C}{z^{3}}
\end{eqnarray}
where $C$=7913.24 \AA$^{3}$ K and $D$=259.39 K are parameters derived from the original \paraH2-C Lennard-Jones parameters\cite{levesque02} ($\sigma=3.18$ \AA, $\epsilon=32.05$ K) and the density of carbon atoms in graphite ($\rho=0.114$ \AA$^{-3}$).   

The assumption built in (\ref{hm}) of a smooth graphite substrate seems justified, on account of the relatively large ($\sim$ 6 \AA) distance between the hydrogen molecules and the substrate, due to the presence of the Kr layer. Nevertheless, corrugation may be important in this problem, given that the Kr layer is not registered with the underlying graphite substrate. Therefore, in order to gauge the effect of substrate corrugation, we  performed a number of simulations with an explicitly modeled top layer of C atoms, placing a smooth graphite slab underneath. For these simulations, the pair-wise interaction between a hydrogen molecule and a C atom was taken to be the same LJ potential utilized to infer the ``3-9" interaction of hydrogen molecules with a smooth substrate.

The results obtained using this composite substrate indicate no qualitative change in the physical character of the system as a result of introducing the smooth plane approximation for graphite.  Quantitatively, the potential energy per particle, for example, was found to be $\sim$2 K lower when assuming explicit corrugation of the graphite substrate, a difference of less than 1\%. Henceforth, therefore, we shall confine our discussion to the case of a smooth graphite substrate, i.e., model (\ref{hm}).

\begin{figure}            %
\centerline{\includegraphics[scale=0.6]{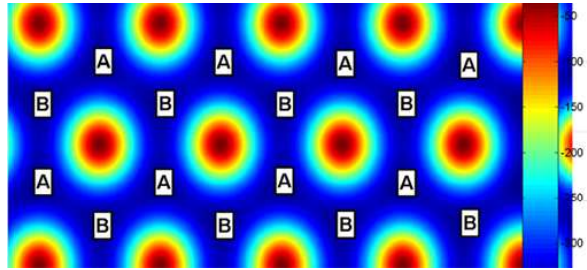}}
\caption{(color online) Potential energy surface, at the vertical equilibrium distance of molecular hydrogen, as a function of position.  The two energetically degenerate sublattices of preferential adsorption sites, A and B, are shown.  A \textit{commensurate} coverage is defined as the occupation of one entire sublattice (also termed ``$\frac{1}{2}$ filling'').  This is a low resolution image; contact the authors for high resolution image.}  
\label{sublattice}
\end{figure}

A slice of the effective potential energy surface for molecular hydrogen resting
at its average height above the substrate is shown in Figure \ref{sublattice}; included are the interactions
with both the Kr monolayer and the graphite slab.  There are two energetically degenerate sublattices of
preferential adsorption sites for this geometry - these are denoted sublattice \fbox{A} and sublattice \fbox{B}.

The model (\ref{hm}) clearly contains important physical simplifications, such as the neglect of zero-point motion of Kr atoms, as well as the restriction to additive pairwise interactions (to the exclusion of, for example, three-body terms), all taken to be central, and the use of the highly simplified LJ and ``3-9" potentials. Nonetheless, it seems a reasonable starting point, and even quantitatively we expect it to capture the bulk of the physical picture.

\section{COMPUTATIONAL METHOD}  
\label{method}

Accurate ground state expectation values for quantum many-body systems
described by a Hamiltonian such as (\ref{hm}) can be computed 
by means of Quantum Monte Carlo (QMC) simulations. 
In this work, the method utilized is \textit{Path Integral ground state} 
(PIGS), which is an extension to zero temperature of the standard, 
Path Integral Monte Carlo method.\cite{Ceperley1}  
PIGS is a \textit{projection technique}, which filters the exact ground state wave 
function out of an initial trial state. It is therefore closely related to
other ground state projection methods, such as Diffusion Monte Carlo (DMC),
but has a few distinct advantages (for a discussion, see, for instance, Ref. 
\onlinecite{Sarsa1}).  
Because this method is described in a number of publications, it will not be reviewed here. Some of the technical details of the calculation performed in this work (mainly, the short imaginary time propagator) are the same as in Ref. \onlinecite{boninsegni04a}.

The trial wave function utilized is of the Jastrow type:
\begin{eqnarray}\label{trial}
\Psi_T({\bf r}_1,{\bf r}_2,...{\bf r}_N)&=& \biggl ( \prod_{i<j}^N e^{-v(r_{ij})}\biggr ) \times \biggl ( \prod_{i=1}^N e^{-w(z_i)}\biggr ) \nonumber \\
&\times& \biggl (\prod_{i=1}^N\prod_{k=1}^L e^{-u(|{\bf r}_i-{\bf R}_k|)}\biggr ) 
\end{eqnarray}
with pseudo-potentials $w$ (\paraH2-graphite and \orthoD2), $u$ (\paraH2-Kr and \orthoD2-Kr), and $v$ (\paraH2-\paraH2 and \orthoD2-\orthoD2) chosen as follows:
\begin{equation}
w(r)=\frac{\alpha}{z^3} \; , \;\; u(r)=\frac{\gamma}{r^5} \;\; {\rm and}  \;\; v(r)=\frac{\mu}{r^5}
\end{equation}

The values of the parameters $\alpha=30$ \AA$^{3}$, $\gamma=250$ 
\AA$^{5}$ and $\mu=750$ \AA$^{5}$  were obtained empirically, by minimizing the energy 
expectation value computed in separate variational calculations.  Using the trial wavefunction as defined above, 
we observe convergence of the ground state energy 
estimates with a projection time 0.250 K$^{-1}$, using a time step $\tau = (\nu/2560) $ K$^{-1}$, where
$\nu=1$ for \paraH2 and $\nu=2$ for \orthoD2.  

Finite-temperature results were obtained making use of a reformulation of standard path integral Monte Carlo (PIMC) called the \textit{worm algorithm} (WA), a scheme which 
allows for efficient computation of thermodynamic properties, including winding
numbers and off-diagonal correlations, for systems of much greater size than that accessible to
conventional PIMC simulations.\cite{boninsegni06worm1,boninsegni06worm2}
The only input parameter to the WA, as for any finite temperature method, is the microscopic model (\ref{hm}). Just as in ground state  simulations, an optimal value $\tau_{w}$ of the imaginary time step must be determined, such as accurate and unbiased estimates of the observables are obtained; for this system, satisfactory results are yielded by the choice $\tau_w = (\nu$/320) K$^{-1}$.

Calculations for a range of \paraH2 and \orthoD2 coverages were carried out, starting from an initial configuration of 
 molecules sitting atop the Kr layer.  The simulation cell consists of a 12$\times$14  triangular lattice of
Kr atoms with 3.99 \AA\ nearest neighbor spacing (resulting in a simulation box of dimension 47.880 \AA\ $\times$ 48.3762 \AA).\cite{wiechert04} Periodic boundary conditions are used in the three directions, but the simulation cell is chosen sufficiently large in the $z$ direction that they have no effect vertically, because of the strongly attractive character of the composite substrate (i.e., \paraH2 and \orthoD2 molecules do not evaporate).  It is important to note that to capture the features observed experimentally in Ref. \onlinecite{wiechert04} requires that a large number of particles be simulated (the simulation cell must be large enough to support at least two crystalline domains).  

The systematic errors of our calculation are attributable to finite
projection time (for the $T$=0 PIGS calculations) and the finite time step $\tau$ (for both PIGS and the WA).  
Based on comparisons of results obtained from simulations with different 
values of projection time and/or time step, we estimate our combined systematic error on the 
total energy per \paraH2 molecule to be of the order of 0.7 K or less (corresponding to less 0.5\%).  The simultaneous use of a ground state and a finite temperature method allows us to obtain an independent check of our calculations in the $T\to 0$ limit, where the two methods must yield the same results, within statistical uncertainties.

\section{RESULTS}  
\label{results}

Physical quantities of primary interest, for both \paraH2 and \orthoD2, include the energy per molecule, $e(\theta)$, and the superfluid fraction, $\rho_s$.  PIGS ($T$=0) was used to compute $e(\theta)$ as a function of the coverage (two-dimensional density), $\theta$, from $\theta$=0.06735 \AA$^{-2}$ to $\theta$=0.09498 \AA$^{-2}$.  For a subset of these coverages, the worm algorithm (WA) was employed to yield finite temperature estimates for all of the above physical quantities, for a set of temperatures between $T$=5 K and $T$=0.25 K.
 
\begin{figure} [h]           
\centerline{\includegraphics[scale=0.7]{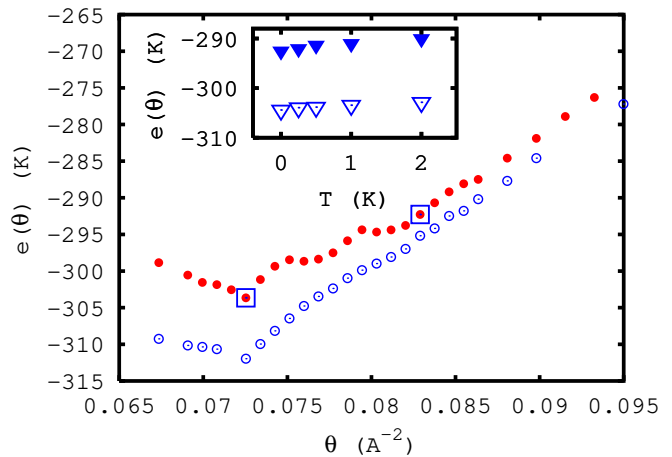}}
\caption{(color online) Energy per molecule $e(\theta)$ (in K)  for \paraH2 ($\circ$), shifted by $-50$ K for clarity, and \orthoD2 ($\bullet$) computed by PIGS ($T$=0), as a function of the coverage $\theta$ (in \AA$^{-2}$).  Inset: $e(\theta)$ calculated using WA for \orthoD2 is shown with $\blacktriangledown$ for $\theta$=0.08389 \AA$^{-2}$ and $\triangledown$ for $\theta$=0.07253 \AA$^{-2}$, values marked with $\Box$ on the $T$=0 \orthoD2 curve.}  
\label{energyplot}
\end{figure}

The results for $e(\theta)$ are shown in Figure \ref{energyplot}; note that $e(\theta)$ for \paraH2 has been shifted by $-50$ K for clarity. 
In the case of \paraH2, the main feature is an energetic minimum at $\theta$=0.07253 \AA$^{-2}$, corresponding to the formation of a thermodynamically stable solid layer, that is commensurate with the underlying krypton lattice.  This coverage is approximately 8\% greater than the equilibrium density of \paraH2 in strictly 2D.\cite{boninsegni04b}   Within the precision of the calculation, $e(\theta)$ is numerically consistent with the absence of other thermodynamically stable coverages, in the range explored in this work.

\begin{figure}           
\centerline{\includegraphics[scale=0.7]{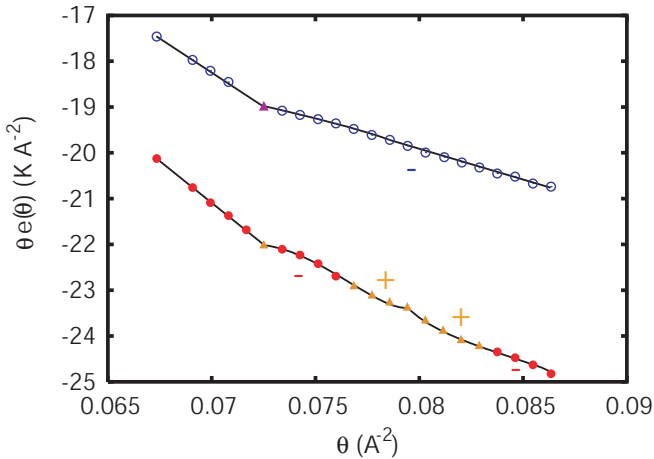}}
\caption{(color online) The product $\theta e(\theta)$ (in K\AA$^{-2}$) for \paraH2 ($\circ$), and for \orthoD2 ($\bullet$) computed by PIGS ($T$=0), as a function of the coverage $\theta$ (in \AA$^{-2}$).  $\blacktriangle$ denote coverages within regions of positive curvature.}  
\label{comp_plot}
\end{figure}

For \orthoD2, we find that there is an energetic minimum at $\theta$=0.07253 \AA$^{-2}$, again corresponding to a commensurate solid film.  In addition, one finds two coverage ranges, separated by a cusp, where $\theta e(\theta)$, shown in Figure \ref{comp_plot}, has a positive second derivative (meaning that these configurations satisfy the condition of positive compressibility, ${d\mu}/{d\theta}$=${d^{2}}[\theta e(\theta)]/{d\theta^{2}}>0$, and are thus thermodynamically stable); we will call the first of these IC-I, and the second IC-II.  IC-I extends from $\theta$=0.07598 \AA$^{-2}$ to $\theta$=0.08030 \AA$^{-2}$, while IC-II extends from $\theta$=0.08030 \AA$^{-2}$ to $\theta$=0.8635 \AA$^{-2}$.  The much richer behavior of $e(\theta)$ in the case of \orthoD2, is a direct consequence of  the greater mass, and reduced zero-point motion, of this isotope; as a result, the potential energy plays a greater role in shaping the physical properties.  Phases of intermediate coverages found for \orthoD2 are apparently ``washed out" by zero-point motion in the case of \paraH2.  While this result is surprising, similar anomalous behavior is noted in related studies.\cite{Wiechert86,Wiechert03}

\begin{figure}            
\centerline{\includegraphics[height=2.65in]{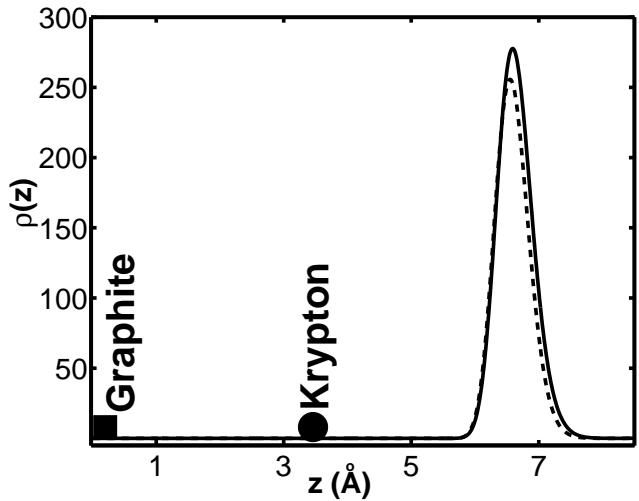}}
\caption{(color online) Density profile of \orthoD2 adsorbed to the composite substrate for $\theta$=0.08389 \AA$^{-2}$ and  $\theta$=0.07253 \AA$^{-2}$ at $T$=1 K (solid line and dashed line respectively).  The square represents the position of graphite (at $z$=0 \AA) and the circle represents the position of the Kr layer (at $z$=3.46 \AA).}   
\label{profile}
\end{figure} 

Of particular interest for the present study is IC-II, where the \orthoD2 is found to form a structure best characterized as being composed of several crystalline domains commensurately occupying one of the two triangular sublattices available.  Between these commensurate domains are regions of high \orthoD2 density - domain walls.  Molecules that are part of these domain walls have a decreased average interparticle distance.  One may expect that such \orthoD2, having the same zero-point motion as \orthoD2 in the commensurate domains, would have a greater likelihood of quantum exchange, in particular if there is domain-wall disorder.  Indeed, IC-II includes precisely the coverages reported in the experiments of Ref. \onlinecite{wiechert04} to exhibit anomalous fluid-like signatures down to $T$=1.5 K, well below the expected freezing point.  

In order to provide accurate permutation statistics and superfluid density estimates, additional simulations were carried out for several coverages at finite temperature using the worm algorithm.  Figure \ref{energyplot}, inset, shows the energy values obtained in the range of temperature 0.25 K $\le$ T $\le$ 5 K,  for $\theta$=0.07253 \Am2 and $\theta$=0.08389 \Am2.  

Shown in Figure \ref{profile} are vertical density profiles, $\rho(z)$, for $\theta$=0.07253 \AA$^{-2}$ and $\theta$=0.08389 \AA$^{-2}$ at $T$=1 K.  One can see that the vertical extent of the \orthoD2 film is slightly enhanced for increased coverages.  

\begin{figure}             
\centering
\def\subfigcapskip{-17pt}
\subfigure[\paraH2 with $\theta$=0.07253 \AA$^{-2}$ at $T$=0 K.]{\includegraphics[height=2.8in]{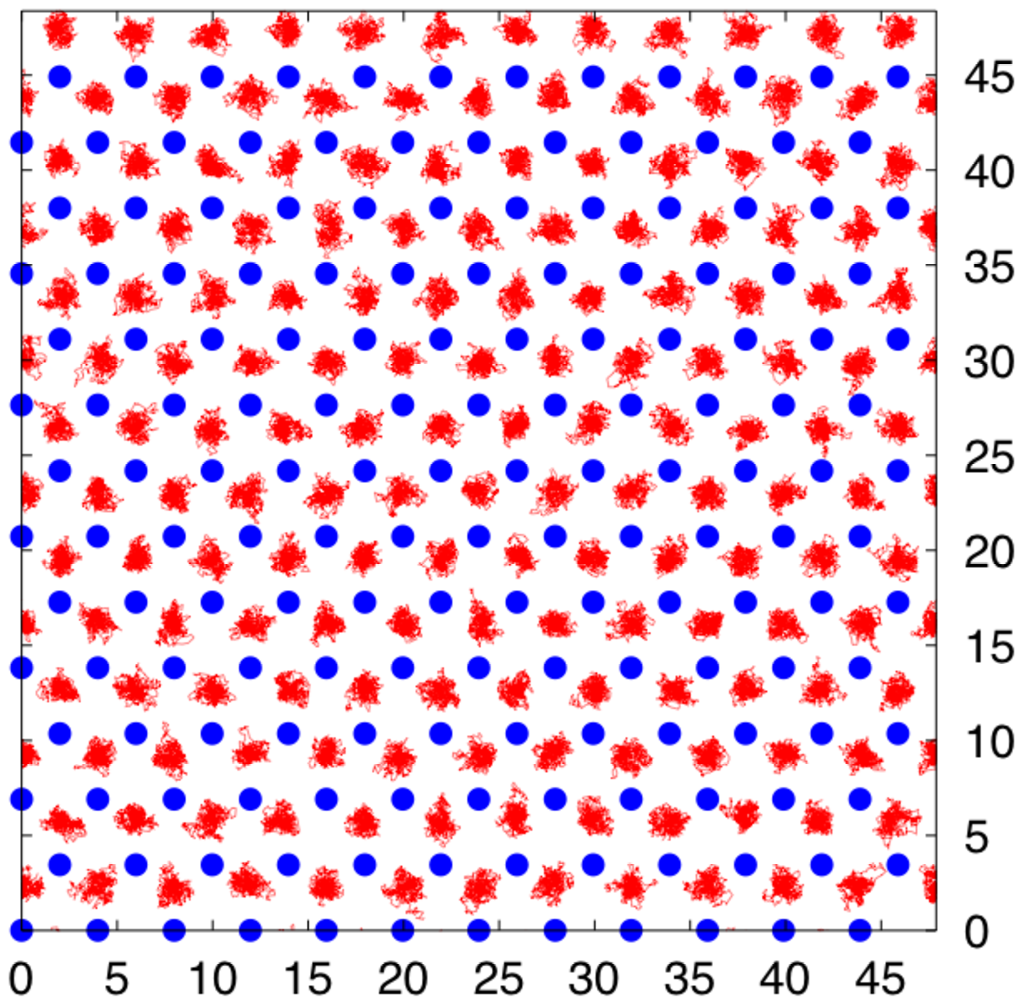}}{\label{snap1}} \vspace{-0.2in} \\

\def\subfigcapskip{-17pt}
\subfigure[\orthoD2 with $\theta$=0.07253 \AA$^{-2}$ at $T$=0 K.]{\includegraphics[height=2.8in]{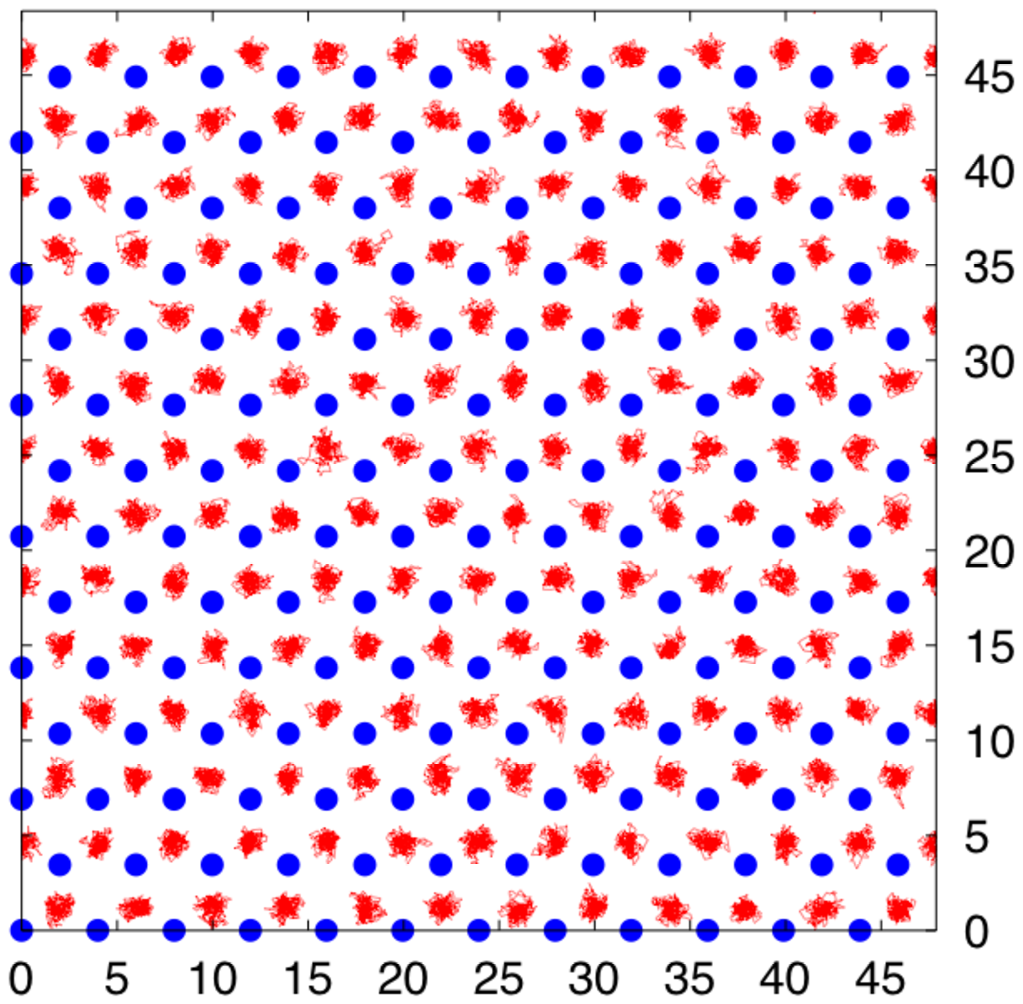}}{\label{snap2}} \vspace{-0.2in} \\

\def\subfigcapskip{-17pt}
\subfigure[\orthoD2 with $\theta$=0.08389 \AA$^{-2}$ at $T$=0.5 K.]{\includegraphics[height=2.8in]{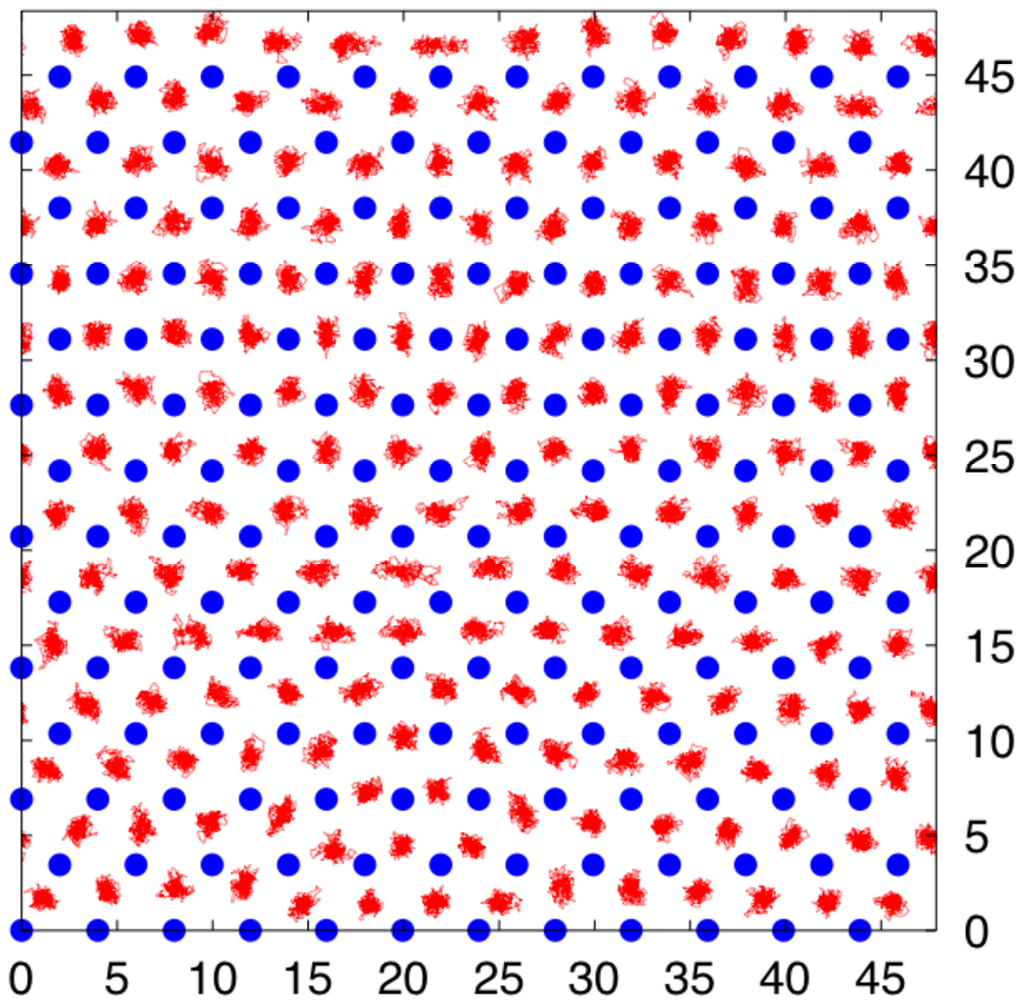}}{\label{snap3}} \vspace{-0.1in} \\

\caption{(color online) Snapshots of typical configurations of \paraH2 molecules adsorbed to the graphite/compressed Kr substrate.  The positions of all molecules at each one of the imaginary time slices are shown as discrete paths. Distances are expressed in \AA.  These are low resolution images; contact the authors for high resolution images.}

\label{snapshots}
\end{figure}

While dynamical information cannot be extracted using the computational techniques employed in this study, one does have access to real-space instantaneous configurations of the system, from which one can extract qualitative insight, and potentially use to preclude particular system behaviors.   Such representative many-particle configurations are shown in Figure \ref{snapshots} for \orthoD2 and \paraH2 with $\theta$=0.07253 \Am2 at $T=0$, and for \orthoD2 with $\theta$=0.08389 \Am2 at $T$=0.5 K.  In each of (a) to (c), we find a high degree of particle localization, even along the domain walls observed in (c).  No single configuration observed was found to be inconsistent with the instantaneous configurations one would necessarily observe in a domain-wall fluid, while the spatial configuration of domain-walls was not fixed as the system evolved in Monte Carlo time.  

Finally, consistent with the high degree of \orthoD2 localization observed, permutations are found to not take place at any temperature for any coverage examined, and, accordingly, the finite temperature estimator for the superfluid density consistently yields a value of $\rho_s$=0.  In light of the absence of structural and energetic differences found between $T$=0 and $T$=0.25 K, it is unlikely that novel phases of \orthoD2 (namely, phases with a finite superfluid density) remain unexplored at intermediate temperatures.

\section{DISCUSSION}
\label{conclusions}

Using numerically exact ground state Quantum Monte Carlo (PIGS) and finite-temperature path integral Monte Carlo (worm algorithm) methods, we studied \paraH2 and \orthoD2 adsorption onto a graphite substrate preplated with a compressed krypton monolayer.  We performed calculations based a simple model, in which graphite corrugation is ignored, the Kr atoms in the spacer layer are assumed static and point-like, and \paraH2(\orthoD2)-substrate interactions are given by Lennard-Jones type potentials. 
Using an exact ground state technique, for both \paraH2 and \orthoD2, we find that a solid monolayer commensurate with the Kr layer is thermodynamically stable, and, in both cases, is the coverage corresponding to the minimum energy per particle.  For \orthoD2, we also find that there are two distinct compressible incommensurate solid regions, with the denser region, IC-II, corresponding to regions of commensuration separated by domain walls.  Investigation of this coverage region at finite temperature, using the worm algorithm, down to $T$=0.25 K, yields no evidence of superfluidity, consistent with our observation of a high degree of localization of \orthoD2 molecules.  No quantum exchanges, and by extension permutations, are found to take place, though they are sampled efficiently.
Though our simulations support  the notion of a ``domain-wall fluid'', as all \textit{instantaneous configurations} generated by our QMC algorithm do indeed show such domain walls, this mechanism is found not to  support superfluidity of either hydrogen isotope.  

There are obviously several sources of uncertainty in this calculation which need to be discussed.  The potentials used to describe the interactions between the \paraH2 and the substrate are very rough; this does not seem too important an issue, as far as the interaction of \paraH2 molecules with graphite is concerned, given the relatively large average distance at which molecules sit, supported by calculations using explicit corrugation of the graphite substrate. On the other hand, a more realistic interaction potential between \paraH2 and krypton may \textit{quantitatively} alter the energetics shown here.  Despite these issues, and other simplifications, it does not seem likely that the qualitative structural information will change.  
Thus, we conclude this system is not a good candidate for further consideration in the search for superfluid \orthoD2 or \paraH2.

In general, the results shown here confirm those obtained in a other studies,\cite{njp} to the effect that frustration arising from incommensuration does not lead to the formation of a stable superfluid phase of molecular hydrogen at low $T$. A more complex physical mechanism, causing a weakening of the effective interaction between two hydrogen molecules may have to be investigated, if some progress is to be achieved in this direction. Possibly,  the interaction of the hydrogen molecules with the surface 
electrons of a metal substrate, or of a nanostructure, might have such a ``renormalizing" effect.
\section*{Acknowledgments}
J. T. would like to thank Horst Wiechert (Johannes-Gutenberg-Universitat) for fruitful discussion and insight.  M. B. acknowledges kind hospitality of the Theoretical Physics Institute, ETH, Z\"urich. This work was supported by the Natural Sciences and Engineering Research council of Canada (NSERC) under research grant G121210893, and by the Alberta Informatics Circle of Research Excellence (ICORE). Computing support from Westgrid and from AICT, University of Alberta, is gratefully acknowledged.


\begin{thebibliography}{34}

\expandafter\ifx\csname natexlab\endcsname\relax\def\natexlab#1{#1}\fi
\expandafter\ifx\csname bibnamefont\endcsname\relax
  \def\bibnamefont#1{#1}\fi
\expandafter\ifx\csname bibfnamefont\endcsname\relax
  \def\bibfnamefont#1{#1}\fi
\expandafter\ifx\csname citenamefont\endcsname\relax
  \def\citenamefont#1{#1}\fi
\expandafter\ifx\csname url\endcsname\relax
  \def\url#1{\texttt{#1}}\fi
\expandafter\ifx\csname urlprefix\endcsname\relax\def\urlprefix{URL }\fi
\providecommand{\bibinfo}[2]{#2}
\providecommand{\eprint}[2][]{\url{#2}}



\bibitem[{\citenamefont{Ginzburg and Sobyanin}(1972)}]{ginzburg72}
\bibinfo{author}{\bibfnamefont{V.~L.} \bibnamefont{Ginzburg}} \bibnamefont{and}
  \bibinfo{author}{\bibfnamefont{A.~A.} \bibnamefont{Sobyanin}},
  \bibinfo{journal}{JETP} \textbf{\bibinfo{volume}{15}}, \bibinfo{pages}{242}
  (\bibinfo{year}{1972}).

\bibitem[{\citenamefont{Bretz and Thomson}(1981)}]{bretz81}
\bibinfo{author}{\bibfnamefont{M.}~\bibnamefont{Bretz}} \bibnamefont{and}
  \bibinfo{author}{\bibfnamefont{A.~L.} \bibnamefont{Thomson}},
  \bibinfo{journal}{Phys. Rev. B} \textbf{\bibinfo{volume}{24}},
  \bibinfo{pages}{467} (\bibinfo{year}{1981}).

\bibitem[{\citenamefont{Seidel et~al.}(1986)\citenamefont{Seidel, Maris,
  Williams, and Cardon}}]{maris86}
\bibinfo{author}{\bibfnamefont{G.~M.} \bibnamefont{Seidel}},
  \bibinfo{author}{\bibfnamefont{H.~J.} \bibnamefont{Maris}},
  \bibinfo{author}{\bibfnamefont{F.~I.~B.} \bibnamefont{Williams}},
  \bibnamefont{and} \bibinfo{author}{\bibfnamefont{J.~G.}
  \bibnamefont{Cardon}}, \bibinfo{journal}{Phys. Rev. Lett.}
  \textbf{\bibinfo{volume}{56}}, \bibinfo{pages}{2380} (\bibinfo{year}{1986}).

\bibitem[{\citenamefont{Maris et~al.}(1987)\citenamefont{Maris, Seidel, and
  Williams}}]{maris87}
\bibinfo{author}{\bibfnamefont{H.~J.} \bibnamefont{Maris}},
  \bibinfo{author}{\bibfnamefont{G.~M.} \bibnamefont{Seidel}},
  \bibnamefont{and} \bibinfo{author}{\bibfnamefont{F.~I.~B.}
  \bibnamefont{Williams}}, \bibinfo{journal}{Phys. Rev. B}
  \textbf{\bibinfo{volume}{36}}, \bibinfo{pages}{6799} (\bibinfo{year}{1987}).

\bibitem[{\citenamefont{Schindler et~al.}(1996)\citenamefont{Schindler,
  Dertinger, Kondo, and Pobell}}]{schindler96}
\bibinfo{author}{\bibfnamefont{M.}~\bibnamefont{Schindler}},
  \bibinfo{author}{\bibfnamefont{A.}~\bibnamefont{Dertinger}},
  \bibinfo{author}{\bibfnamefont{Y.}~\bibnamefont{Kondo}}, \bibnamefont{and}
  \bibinfo{author}{\bibfnamefont{F.}~\bibnamefont{Pobell}},
  \bibinfo{journal}{Phys. Rev. B} \textbf{\bibinfo{volume}{53}},
  \bibinfo{pages}{11451} (\bibinfo{year}{1996}).

\bibitem[{\citenamefont{Boninsegni}(2004{\natexlab{a}})}]{boninsegni04b}
\bibinfo{author}{\bibfnamefont{M.}~\bibnamefont{Boninsegni}},
  \bibinfo{journal}{Phys. Rev. B} \textbf{\bibinfo{volume}{70}},
  \bibinfo{pages}{193411} (\bibinfo{year}{2004}{\natexlab{a}}).

\bibitem[{\citenamefont{Boninsegni et~al.}(1999)\citenamefont{Boninsegni, Cole,
  and Toigo}}]{boninsegni99}
\bibinfo{author}{\bibfnamefont{M.}~\bibnamefont{Boninsegni}},
  \bibinfo{author}{\bibfnamefont{M.~W.} \bibnamefont{Cole}}, \bibnamefont{and}
  \bibinfo{author}{\bibfnamefont{F.}~\bibnamefont{Toigo}},
  \bibinfo{journal}{Phys. Rev. Lett.} \textbf{\bibinfo{volume}{83}},
  \bibinfo{pages}{2002} (\bibinfo{year}{1999}).

\bibitem[{\citenamefont{Nielsen et~al.}(1980)\citenamefont{Nielsen, McTague,
  and Passell}}]{nielsen80}
\bibinfo{author}{\bibfnamefont{M.}~\bibnamefont{Nielsen}},
  \bibinfo{author}{\bibfnamefont{J.~P.} \bibnamefont{McTague}},
  \bibnamefont{and} \bibinfo{author}{\bibfnamefont{L.}~\bibnamefont{Passell}},
  in \emph{\bibinfo{booktitle}{Phase Transitions in Surface Films}}, edited by
  \bibinfo{editor}{\bibfnamefont{J.}~\bibnamefont{Dash}} \bibnamefont{and}
  \bibinfo{editor}{\bibfnamefont{J.}~\bibnamefont{Ruvalds}}
  (\bibinfo{publisher}{Plenum}, \bibinfo{address}{New York},
  \bibinfo{year}{1980}).

\bibitem[{\citenamefont{Lauter et~al.}(1990)\citenamefont{Lauter, Godfrin,
  Frank, and Leiderer}}]{lauter90}
\bibinfo{author}{\bibfnamefont{H.~J.} \bibnamefont{Lauter}},
  \bibinfo{author}{\bibfnamefont{H.}~\bibnamefont{Godfrin}},
  \bibinfo{author}{\bibfnamefont{V.~L.~P.} \bibnamefont{Frank}},
  \bibnamefont{and} \bibinfo{author}{\bibfnamefont{P.}~\bibnamefont{Leiderer}},
  in \emph{\bibinfo{booktitle}{Phase Transitions in Surface Films 2}}, edited
  by \bibinfo{editor}{\bibfnamefont{H.~J.~L.} \bibnamefont{H.~Taub},
  \bibfnamefont{G.~Torzo}} \bibnamefont{and}
  \bibinfo{editor}{\bibfnamefont{S.~C.~F.} \bibnamefont{Jr.}}
  (\bibinfo{publisher}{Plenum}, \bibinfo{address}{New York},
  \bibinfo{year}{1990}).

\bibitem[{\citenamefont{Wiechert}(1991)}]{wiechert91}
\bibinfo{author}{\bibfnamefont{H.}~\bibnamefont{Wiechert}}, in
  \emph{\bibinfo{booktitle}{Excitations in Two-Dimensional and
  Three-Dimensional Quantum Fluids}}, edited by
  \bibinfo{editor}{\bibfnamefont{A.~F.~G.} \bibnamefont{Wyatt}}
  \bibnamefont{and} \bibinfo{editor}{\bibfnamefont{H.~J.} \bibnamefont{Lauter}}
  (\bibinfo{publisher}{Plenum}, \bibinfo{address}{New York},
  \bibinfo{year}{1991}).

\bibitem[{\citenamefont{Liu et~al.}(1992)\citenamefont{Liu, Liu, and
  Vilches}}]{vilches92}
\bibinfo{author}{\bibfnamefont{F.~C.} \bibnamefont{Liu}},
  \bibinfo{author}{\bibfnamefont{Y.~M.} \bibnamefont{Liu}}, \bibnamefont{and}
  \bibinfo{author}{\bibfnamefont{O.~E.} \bibnamefont{Vilches}},
  \bibinfo{journal}{J. Low Temp. Phys.} \textbf{\bibinfo{volume}{89}},
  \bibinfo{pages}{649} (\bibinfo{year}{1992}).

\bibitem[{\citenamefont{Cheng et~al.}(1993)\citenamefont{Cheng, Mistura, Lee,
  Chan, Cole, Carraro, Saam, and Toigo}}]{cheng93b}
\bibinfo{author}{\bibfnamefont{E.}~\bibnamefont{Cheng}},
  \bibinfo{author}{\bibfnamefont{G.}~\bibnamefont{Mistura}},
  \bibinfo{author}{\bibfnamefont{H.~C.} \bibnamefont{Lee}},
  \bibinfo{author}{\bibfnamefont{M.~H.~W.} \bibnamefont{Chan}},
  \bibinfo{author}{\bibfnamefont{M.~W.} \bibnamefont{Cole}},
  \bibinfo{author}{\bibfnamefont{C.}~\bibnamefont{Carraro}},
  \bibinfo{author}{\bibfnamefont{W.~F.} \bibnamefont{Saam}}, \bibnamefont{and}
  \bibinfo{author}{\bibfnamefont{F.}~\bibnamefont{Toigo}},
  \bibinfo{journal}{Phys. Rev. Lett.} \textbf{\bibinfo{volume}{70}},
  \bibinfo{pages}{1854} (\bibinfo{year}{1993}).

\bibitem[{\citenamefont{Mistura et~al.}(1994)\citenamefont{Mistura, Lee, and
  Chan}}]{mistura94}
\bibinfo{author}{\bibfnamefont{G.}~\bibnamefont{Mistura}},
  \bibinfo{author}{\bibfnamefont{H.~C.} \bibnamefont{Lee}}, \bibnamefont{and}
  \bibinfo{author}{\bibfnamefont{M.~H.~W.} \bibnamefont{Chan}},
  \bibinfo{journal}{J. Low Temp. Phys.} \textbf{\bibinfo{volume}{96}},
  \bibinfo{pages}{221} (\bibinfo{year}{1994}).

\bibitem[{\citenamefont{Ross et~al.}(1998)\citenamefont{Ross, Taborek, and
  Rutledge}}]{ross98}
\bibinfo{author}{\bibfnamefont{P.}~\bibnamefont{Ross}},
  \bibinfo{author}{\bibfnamefont{P.}~\bibnamefont{Taborek}}, \bibnamefont{and}
  \bibinfo{author}{\bibfnamefont{J.~E.} \bibnamefont{Rutledge}},
  \bibinfo{journal}{Phys. Rev. B} \textbf{\bibinfo{volume}{58}},
  \bibinfo{pages}{4274} (\bibinfo{year}{1998}).

\bibitem[{\citenamefont{Wiechert et~al.}(2004)\citenamefont{Wiechert, Kortmann,
  and St$\ddot{u}$ber}}]{wiechert04}
\bibinfo{author}{\bibfnamefont{H.}~\bibnamefont{Wiechert}},
  \bibinfo{author}{\bibfnamefont{K.}~\bibnamefont{Kortmann}}, \bibnamefont{and}
  \bibinfo{author}{\bibfnamefont{N.}~\bibnamefont{St$\ddot{u}$ber}},
  \bibinfo{journal}{Phys. Rev. B} \textbf{\bibinfo{volume}{70}},
  \bibinfo{pages}{125410} (\bibinfo{year}{2004}).

\bibitem[{\citenamefont{Turnbull and Boninsegni}(2005)}]{turnbull05}
\bibinfo{author}{\bibfnamefont{J.~D.} \bibnamefont{Turnbull}} \bibnamefont{and}
  \bibinfo{author}{\bibfnamefont{M.}~\bibnamefont{Boninsegni}},
  \bibinfo{journal}{J. Low Temp. Phys.} \textbf{\bibinfo{volume}{140}},
  \bibinfo{pages}{269} (\bibinfo{year}{2005}).

\bibitem[{\citenamefont{Boninsegni}(2004{\natexlab{b}})}]{boninsegni04_li}
\bibinfo{author}{\bibfnamefont{M.}~\bibnamefont{Boninsegni}},
  \bibinfo{journal}{Phys. Rev. B} \textbf{\bibinfo{volume}{70}},
  \bibinfo{pages}{125405} (\bibinfo{year}{2004}{\natexlab{b}}).

\bibitem[{\citenamefont{Boninsegni
  et~al.}(2006{\natexlab{a}})\citenamefont{Boninsegni, Prokofev, and
  Svistunov}}]{boninsegni06worm1}
\bibinfo{author}{\bibfnamefont{M.}~\bibnamefont{Boninsegni}},
  \bibinfo{author}{\bibfnamefont{N.~V.} \bibnamefont{Prokofev}},
  \bibnamefont{and} \bibinfo{author}{\bibfnamefont{B.~V.}
  \bibnamefont{Svistunov}}, \bibinfo{journal}{Phys. Rev. Lett.}
  \textbf{\bibinfo{volume}{96}}, \bibinfo{pages}{070601}
  (\bibinfo{year}{2006}{\natexlab{a}}).

\bibitem[{\citenamefont{Boninsegni
  et~al.}(2006{\natexlab{b}})\citenamefont{Boninsegni, Prokofev, and
  Svistunov}}]{boninsegni06worm2}
\bibinfo{author}{\bibfnamefont{M.}~\bibnamefont{Boninsegni}},
  \bibinfo{author}{\bibfnamefont{N.~V.} \bibnamefont{Prokofev}},
  \bibnamefont{and} \bibinfo{author}{\bibfnamefont{B.~V.}
  \bibnamefont{Svistunov}}, \bibinfo{journal}{Phys. Rev. E}
  \textbf{\bibinfo{volume}{74}}, \bibinfo{pages}{036701}
  (\bibinfo{year}{2006}{\natexlab{b}}).

\bibitem[{\citenamefont{Steele}(1973)}]{steele}
\bibinfo{author}{\bibfnamefont{W.~A.} \bibnamefont{Steele}},
  \bibinfo{journal}{Surf. Sci.} \textbf{\bibinfo{volume}{36}},
  \bibinfo{pages}{317} (\bibinfo{year}{1973}).

\bibitem[{\citenamefont{Gooding et~al.}(1983)\citenamefont{Gooding, Joos, and
  Bergersen}}]{gooding}
\bibinfo{author}{\bibfnamefont{R.~J.} \bibnamefont{Gooding}},
  \bibinfo{author}{\bibfnamefont{B.}~\bibnamefont{Joos}}, \bibnamefont{and}
  \bibinfo{author}{\bibfnamefont{B.}~\bibnamefont{Bergersen}},
  \bibinfo{journal}{Phys. Rev. B} \textbf{\bibinfo{volume}{27}},
  \bibinfo{pages}{7669} (\bibinfo{year}{1983}).

\bibitem[{\citenamefont{Silvera and Goldman}(1978)}]{Silvera1}
\bibinfo{author}{\bibfnamefont{I.~F.} \bibnamefont{Silvera}} \bibnamefont{and}
  \bibinfo{author}{\bibfnamefont{V.~V.} \bibnamefont{Goldman}},
  \bibinfo{journal}{J. Chem. Phys.} \textbf{\bibinfo{volume}{69}},
  \bibinfo{pages}{4209} (\bibinfo{year}{1978}).

\bibitem[{\citenamefont{Wang et~al.}(1996)\citenamefont{Wang, Johnson, and
  Broughton}}]{johnson96}
\bibinfo{author}{\bibfnamefont{Q.}~\bibnamefont{Wang}},
  \bibinfo{author}{\bibfnamefont{J.~K.} \bibnamefont{Johnson}},
  \bibnamefont{and} \bibinfo{author}{\bibfnamefont{J.~Q.}
  \bibnamefont{Broughton}}, \bibinfo{journal}{Mol. Phys.}
  \textbf{\bibinfo{volume}{89}}, \bibinfo{pages}{1105} (\bibinfo{year}{1996}).

\bibitem[{\citenamefont{Operetto and Pederiva}(2004)}]{operetto}
\bibinfo{author}{\bibfnamefont{F.}~\bibnamefont{Operetto}} \bibnamefont{and}
  \bibinfo{author}{\bibfnamefont{F.}~\bibnamefont{Pederiva}},
  \bibinfo{journal}{Phys. Rev. B} \textbf{\bibinfo{volume}{69}},
  \bibinfo{pages}{024203} (\bibinfo{year}{2004}).

\bibitem[{\citenamefont{Wang and Johnson}(1998)}]{wang97}
\bibinfo{author}{\bibfnamefont{Q.}~\bibnamefont{Wang}} \bibnamefont{and}
  \bibinfo{author}{\bibfnamefont{J.}~\bibnamefont{Johnson}},
  \bibinfo{journal}{Molecular Physics} \textbf{\bibinfo{volume}{95}},
  \bibinfo{pages}{299} (\bibinfo{year}{1998}).

\bibitem[{\citenamefont{MacGowan and Evans}(1986)}]{macgowan86}
\bibinfo{author}{\bibfnamefont{D.}~\bibnamefont{MacGowan}} \bibnamefont{and}
  \bibinfo{author}{\bibfnamefont{D.~J.} \bibnamefont{Evans}},
  \bibinfo{journal}{Phys. Rev. A} \textbf{\bibinfo{volume}{34}},
  \bibinfo{pages}{2133} (\bibinfo{year}{1986}).

\bibitem[{\citenamefont{Le Roy et~al.}(2005)\citenamefont{Le Roy, Wei, Wheatley, and
  Meath}}]{LeRoy05}
\bibinfo{author}{\bibfnamefont{R.~J.} \bibnamefont{Le Roy}},
  \bibinfo{author}{\bibfnamefont{H.}~\bibnamefont{Wei}},
  \bibinfo{author}{\bibfnamefont{R.}~\bibnamefont{Wheatley}}, \bibnamefont{and}
  \bibinfo{author}{\bibfnamefont{W.~J.} \bibnamefont{Meath}},
  \bibinfo{journal}{J. Chem. Phys.} \textbf{\bibinfo{volume}{122}},
  \bibinfo{pages}{084321} (\bibinfo{year}{2005}).

\bibitem[{\citenamefont{Gatica and Johnson}(2004)}]{gatica04}
\bibinfo{author}{\bibfnamefont{S.}~\bibnamefont{Gatica}} \bibnamefont{and}
  \bibinfo{author}{\bibfnamefont{J.}~\bibnamefont{Johnson}},
  \bibinfo{journal}{J. Phys. Chem. B} \textbf{\bibinfo{volume}{108}},
  \bibinfo{pages}{11704} (\bibinfo{year}{2004}).

\bibitem[{\citenamefont{Levesque et~al.}(2002)\citenamefont{Levesque, Gicquel,
  Darkrim, and Kayiran}}]{levesque02}
\bibinfo{author}{\bibfnamefont{D.}~\bibnamefont{Levesque}},
  \bibinfo{author}{\bibfnamefont{A.}~\bibnamefont{Gicquel}},
  \bibinfo{author}{\bibfnamefont{F.}~\bibnamefont{Darkrim}}, \bibnamefont{and}
  \bibinfo{author}{\bibfnamefont{S.}~\bibnamefont{Kayiran}},
  \bibinfo{journal}{J. Phys. CM} \textbf{\bibinfo{volume}{14}},
  \bibinfo{pages}{9285} (\bibinfo{year}{2002}).

\bibitem[{\citenamefont{Ceperley}(1995)}]{Ceperley1}
\bibinfo{author}{\bibfnamefont{D.}~\bibnamefont{Ceperley}},
  \bibinfo{journal}{Rev. Mod. Phys.} \textbf{\bibinfo{volume}{67}},
  \bibinfo{pages}{279} (\bibinfo{year}{1995}).

\bibitem[{\citenamefont{Sarsa et~al.}(2000)\citenamefont{Sarsa, Schmidt, and
  Magro}}]{Sarsa1}
\bibinfo{author}{\bibfnamefont{A.}~\bibnamefont{Sarsa}},
  \bibinfo{author}{\bibfnamefont{K.~E.} \bibnamefont{Schmidt}},
  \bibnamefont{and} \bibinfo{author}{\bibfnamefont{W.~R.} \bibnamefont{Magro}},
  \bibinfo{journal}{J. Chem. Phys.} \textbf{\bibinfo{volume}{113}},
  \bibinfo{pages}{1366} (\bibinfo{year}{2000}).

\bibitem[{\citenamefont{Cuervo et~al.}(2005)\citenamefont{Cuervo, Roy, and
  Boninsegni}}]{boninsegni04a}
\bibinfo{author}{\bibfnamefont{J.~E.} \bibnamefont{Cuervo}},
  \bibinfo{author}{\bibfnamefont{P.-N.} \bibnamefont{Roy}}, \bibnamefont{and}
  \bibinfo{author}{\bibfnamefont{M.}~\bibnamefont{Boninsegni}},
  \bibinfo{journal}{J. Chem. Phys.} \textbf{\bibinfo{volume}{112}},
  \bibinfo{pages}{114504} (\bibinfo{year}{2005}).

\bibitem[{\citenamefont{Freimuth and Wiechert}(1986)}]{Wiechert86}
\bibinfo{author}{\bibfnamefont{H.}~\bibnamefont{Freimuth}} \bibnamefont{and}
  \bibinfo{author}{\bibfnamefont{H.}~\bibnamefont{Wiechert}},
  \bibinfo{journal}{Surf. Sci.} \textbf{\bibinfo{volume}{178}},
  \bibinfo{pages}{716} (\bibinfo{year}{1986}).

\bibitem[{\citenamefont{Wiechert}(2003)}]{Wiechert03}
\bibinfo{author}{\bibfnamefont{H.}~\bibnamefont{Wiechert}}, in
  \emph{\bibinfo{booktitle}{Adsorption of Molecular Hydrogen Isotopes on
  Graphite and Boron Nitride}} (\bibinfo{publisher}{Springer-Verlag},
  \bibinfo{address}{Berlin}, \bibinfo{year}{2003}).

\bibitem[{\citenamefont{njp}(2005)\citenamefont{njp}}]{njp}
  \bibinfo{author}{\bibfnamefont{M.}~\bibnamefont{Boninsegni}},
  \bibinfo{journal}{New J. Phys.} \textbf{\bibinfo{volume}{7}},
  \bibinfo{pages}{78} (\bibinfo{year}{2005}).
\end{thebibliography}

\end{document}